\documentclass[10pt, conference, letterpaper]{IEEEtran}
\usepackage[papersize={8.5in,11in},left=0.625in,right=0.625in,top=0.75in,bottom=1in]{geometry}
\usepackage{blindtext}
\usepackage{graphicx,balance}
\usepackage{mathtools}
\usepackage{cite}
\usepackage[T1]{fontenc}
\usepackage{newtxmath,newtxtext}
\usepackage{blindtext, graphicx, bbm, amsmath, balance}\usepackage{algorithm}
\usepackage[noend]{algpseudocode}
\usepackage{blindtext, graphicx}
\usepackage {tikz}
\usepackage{caption}
\usepackage{amsmath}
\usepackage{amssymb}
\usepackage{algorithm}
\usepackage{float}
\usepackage{algpseudocode}
\usetikzlibrary {positioning}
\usepackage{tkz-euclide}
\usepackage{algorithm,algcompatible,amsmath}
\usepackage[T1]{fontenc}
\usepackage[utf8x]{inputenc}

\ifCLASSINFOpdf
\else
\fi

\usepackage{amsmath, amssymb, amsfonts}

\newcommand\addtag{\refstepcounter{equation}\tag{\theequation}}
\DeclarePairedDelimiter{\abs}{\lvert}{\rvert}

\begin{document}
%
\title{Performance Comparison of Algorithms for Movie Rating Estimation} 

\author{
        \IEEEauthorblockN{Alper Köse\IEEEauthorrefmark{1}\IEEEauthorrefmark{2}, Can Kanbak\IEEEauthorrefmark{2}, Noyan Evirgen\IEEEauthorrefmark{3}}
        \IEEEauthorblockA{
                \small
                \IEEEauthorrefmark{1}Research Laboratory of Electronics, Massachusetts Institute of Technology \\       
                \IEEEauthorrefmark{2}Department of Electrical Engineering, École Polytechnique Fédérale de Lausanne \\
                \IEEEauthorrefmark{3}Department of Information Technology and Electrical Engineering, ETH Zurich \\
                \vspace{-0.5cm}
        }
    
}

\maketitle

\footnotetext[1]{This work has been accepted to the 2017 IEEE ICMLA.}
\begin{abstract}

In this paper, our goal is to compare performances of three different algorithms to predict the ratings that will be given to movies by potential users where we are given a user-movie rating matrix based on the past observations. To this end, we evaluate User-Based Collaborative Filtering, Iterative Matrix Factorization and Yehuda Koren's Integrated model using neighborhood and factorization where we use root mean square error (RMSE) as the performance evaluation metric. In short, we do not observe significant differences between performances, especially when the complexity increase is considered. We can conclude that Iterative Matrix Factorization performs fairly well despite its simplicity.

\end{abstract}


%
\IEEEpeerreviewmaketitle

\section{Introduction}

Recommender systems seek to predict the rating or order of preference that a user would give to an item. Considering the rise of websites like Youtube, Amazon, Netflix and other multimedia content providers, the interest in recommender systems have been drastically increased because of the need to help people for searching what they may like. Websites try to improve their services given to their users by recommending items such as videos, movies, products which they might be interested in.  For example in 2012, Netflix announced that 75\% of what people watch is coming from their recommender system.

Recommender systems are designed to connect customers with items that they would want to have. In order to find these items, they use the information which are provided by the users. Information might have been provided in advance from various sources including their past purchases or items they reviewed. It is crucial to have an accurate recommender system since there are millions of users searching for the best website they can use. To make a recommendation as accurate as possible, all available and meaningful information in the system must be combined with reasonable weights. In this paper, we have implemented three algorithms to recommend movies: User-Based Collaborative Filtering, Iterative Matrix Factorization and Yehuda Koren's integrated method \cite{koren2008factorization} that uses both neighbor and matrix factor elements.

\section{Related Work}

Breese et al. \cite{breese1998empirical} describe several algorithms designed for collaborative filtering and compare their predictive accuracy in a set of representative problem domains. Pazzani \cite{pazzani1999framework} describes the types of information including content of page and demographics of users to determine recommendations. Given a user's preferences for some items, Pennock et al. \cite{pennock2000collaborative} compute the probability that a user will like new items based on her or his probabilistic personality similarity to other users. Linden et al. \cite{linden2003amazon} compare traditional collaborative filtering, cluster models, and search-based methods with their algorithm called item-to-item collaborative filtering. Wang et al. \cite{wang2006unifying} estimate ratings by fusing predictions from three sources: user based collaborative filtering, item based collaborative filtering and ratings predicted based on data from other but similar users rating other but similar items. Ahn \cite{ahn2008new} presents a new heuristic similarity measure, different than pearson correlation coefficient and cosine similarity, that focuses on improving recommendation performance of collaborative filtering under cold start conditions where only a small number of ratings are available for similarity calculation for each user.

Sarwar et al. \cite{sarwar2002incremental} work on factorization to recommend and propose a technique that has the potential to incrementally build SVD-based models to increase scalability. Tak{\'a}cs et al. \cite{takacs2008investigation} propose several matrix factorization approaches with improved prediction accuracy. Ma et al. \cite{ma2008sorec} propose a factor analysis approach based on probabilistic matrix factorization to solve the data sparsity and poor prediction accuracy problems by employing both users' social network information and rating records. Koren et al. \cite{koren2009matrix} discuss the superiority of matrix factorization models to classic nearest-neighbor techniques for producing recommendations by allowing the incorporation of additional information such as implicit feedback, temporal effects, and confidence levels. Jamali and Ester \cite{jamali2010matrix} employ matrix factorization techniques and incorporate the mechanism of trust propagation into the model. Forbes and Zhu \cite{forbes2011content} describe a simple algorithm for incorporating content information directly into the matrix factorization approach.

Recommender systems often rely on collaborative filtering, where past decisions are analyzed in order to establish connections between users and products. Matrix factorization, which directly profile both users and products, and neighborhood models, which analyze similarities between products or users, are two successful approaches to recommend products. Koren \cite{koren2008factorization} merges neighborhood and factorization models to build a more accurate combined model.

\section{Data Handling}
Our data set consists of 10000 users and 1000 movies in which all users rated some of the movies. In other words, there are 10000000 entries in our matrix with 1176192 of them were rated by users in advance. We use 90 percent of given ratings in the matrix for training and remaining 10 percent to validate. The validation set is used to estimate root mean square error ($RMSE$) and assign parameter values through cross validation. Since we provide three different methods to predict missing ratings, we pre-process the data set depending on the method and thus it is explained while introducing the methods.

\section{Methods}

\subsection{User-Based Collaborative Filtering}

In User-Based Collaborative Filtering, given a user and a movie not yet rated by the user, the aim is to estimate the user's rating for this movie by looking at the ratings for the same movie that were given in the past by similar users. This method requires a community of users that provide ratings and a way to assess the similarity between users. The main idea here is that people with similar tastes in the past will have similar tastes in the future. Based on this idea, given a user $U$ and a movie $M$ not rated by $U$, estimate the rating $r_U(M)$ by tracing the steps below:
\\
1. Find a set of users $N_U$ who liked the same items as $U$ in the past and who have rated $M$\\
2. Aggregate the ratings of $M$ provided by $N_U$ with the weights derived from user similarities

To be able to implement this algorithm, we need to define a metric to compute similarity between users, set the number of neighbors (similar users) considered in $N_U$ and a way to aggregate the ratings of $M$ provided by $N_U$.

First of all, we consider two different similarity metrics called pearson correlation coefficient and cosine similarity. To compute pearson correlation coefficient we use the formula:

\begin{equation}
p(a,b)=\cfrac{\sum_{i=1}^{N}{(r_a(i)-\overline{r_a})(r_b(i)-\overline{r_b})}}{\sqrt{\sum_{i=1}^{N}{(r_a(i)-\overline{r_a})^{2}}}\sqrt{\sum_{i=1}^{N}(r_b(i)-\overline{r_b})^{2}}}
\end{equation}

where $p(a,b)$ is the pearson similarity measure. In equality, $a$ and $b$ are users, set $i$ are the movies rated by both $a$ and $b$, $r_a(i)$ is the rating given by user $a$ to movie $i$ and $\overline{r_a}$ is the average rating of $a$.

To compute cosine similarity we use the formula:

\begin{equation}
c(a,b)=\cfrac{\sum_{i=1}^{N}{r_a(i).r_b(i)}}{\sqrt{\sum_{i=1}^{N}{r_a(i)^{2}}}\sqrt{\sum_{i=1}^{N}r_b(i)^{2}}}
\end{equation}

where $c(a,b)$ is the cosine similarity, $a$ and $b$ are users, set $i$ are the movies rated by both $a$ and $b$ and $r_a(i)$ is the rating given by user $a$ to movie $i$.
In the end, aggregation formula is used to estimate ratings:

\begin{equation}
r_a(m)=\overline{r_a}+\cfrac{\sum_{b\in N(a)}{sim(a,b)(r_b(m)-\overline{r_b})}}{\sum_{b\in N(a)}{|sim(a,b)|}}
\end{equation}

where $sim(a,b)$, the similarity of users $a$ and $b$, is defined by either pearson similarity $p(a,b)$ or cosine similarity $c(a,b)$. $N(a)$ are the neighbors of user $a$ and $m$ is a movie not rated by $a$.

This aggregation function calculates whether the neighbors' ratings for the unseen movie $m$ are higher or lower than their average. Then it combines the rating bias using the similarity as a weight, so that the most similar neighbors will have more effect. Finally, the aggregated neighbors' bias is added to user $a$'s average rating.

In this model, we try both pearson similarity and cosine similarity methods for $5,10,25,50,100$ neighbors respectively, and use our validation set to estimate the performance of our settings. As seen from Fig. 1, cosine similarity measure gives better estimates in terms of $RMSE$ evaluation metric and taking more neighbors after $50$ does not lead to much improvement.

User-Based Collaborative Filtering method has some disadvantages. First, it has a cold start problem, that is to say when a new user joins to system, we do not have enough information to decide new user's similarity to existing ones, which leads to random recommendations. Also, this method cannot predict the ratings of new products since no similar user has rated it before. Finally, sparsity may cause a problem by making the process of finding similar users very hard depending on product number vs user number.

\begin{figure}
  \centering
    \includegraphics[height=0.25\textwidth]{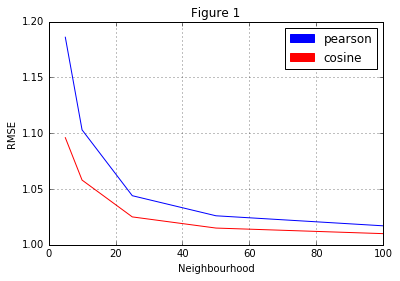}
    \caption{$RMSE$ vs neighborhood size according to similarity measure}
\end{figure}

As a result of this method, we get the best results by using cosine similarity measure and $100$ neighbors for the given user in predictions. $RMSE$ comes out to be $1.01$. Deciding this result is not satisfactory and there is no a room for sufficient improvement, we pass into another method called Iterative Matrix Factorization.

\subsection{Iterative Matrix Factorization}

The main idea of this method is to iteratively construct the low rank approximation of our rating matrix until convergence or a cutoff. This is a simple way to do matrix factorization where each item and user are represented by size $K$ vectors, where $K$ is much smaller than either the number of users or items. These vectors are thought as representing some latent features that explain the given feedback. For example, an element of a movie vector can might represent the 'actionness' of a movie with higher values for action movies and lower values for drama. Similarly, a user with high values for this factor is expected to prefer action movies. So, the rating estimation is simply the inner product of these vectors, which will give high values if the features of the movie aligns with the interests of the user. The matrix composed from these inner products are low-rank, and that's why the low-rank approximation is used for this method.

Firstly, we subtract average rating of each movie, namely each of user's ratings are shifted relative to each movie's average rating. On the other hand, unrated entries in the matrix are given zero. After this operation, we recursively perform low rank approximation via singular value decomposition of the ratings matrix and after each approximation, we put the known values from our training matrix to the same places in low rank approximation matrix. We continue to implement this process until we get the best result of estimation in terms of $RMSE$. In the end, we shift all the entries in matrix by adding movie's means to related columns.

In this method, we have two things to find. First, we need to search for the optimal rank for the reconstruction process and then we should spot the number of iterations which gives the best results. Making use of our validation set, we observe that higher ranks in the approximation do not give better results and by trying lower values, we reach at the results seen in Fig. 2. As can be seen from Fig. 2, prediction results improve until rank $3$ and then decrease. Only first $5$ values of rank is illustrated in the figure for a clear view. We get the best estimations by using rank $3$ in lower order approximations of the ratings matrix and doing the iteration $20$ times. The result corresponds to an $RMSE$ of $0.9908$.

\begin{figure}
  \centering
    \includegraphics[height=0.25\textwidth]{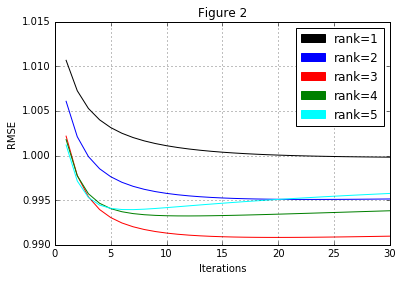}
    \caption{$RMSE$ vs iterations according to rank of lower order approximation}
\end{figure}

In Iterative Matrix Factorization, we end up having improved results compared to User-Based Collaborative Filtering method. However, thinking that this estimation of ratings are not also qualitative enough, we pass into another method which was used in Netflix competition and leaded to successful predictions.

\subsection{The Integrated Model \cite{koren2008factorization}}
\subsubsection{Reasoning}
The last method we implement is the integrated model that is designed by Koren \cite{koren2008factorization}. It is called integrated because it combines both a matrix factorization approach and a neighborhood approach to make the predictions. We have shown two simple models to use these approaches that do not work very well but still able to show the reasoning behind them. This approach combines these methods and accompanying reasons to create an even better recommender system. The prediction formula can be written as:
\begin{align*}
\hat{r}_{ui} &= \mu + b_u + b_i + q_i^T\left(p_u + \abs{N(u)}^{-1/2}\sum_{j\in N(u)}y_j\right) \\
&+\abs{R^k(i,u)}^{-1/2}\sum_{t\in R^k(i;u)}(r_{ut}-b_{ut})w_{it} \\
&+\abs{N^k(i,u)}^{-1/2}\sum_{t\in N^k(i;u)}c_{it}  \addtag \label{1}
\end{align*}
This formula becomes simpler if the elements are analyzed separately. First value $\mu$ is the overall average of the given ratings. It is the empirical expected value and it is what we would output if we knew nothing about the user or the item. $b_u$ is the user baseline value, which is the average rating of the user $u$ minus the total average and $b_i$ is the item baseline - average rating for the item $i$ minus the total average. In total, $b_{ui}=\mu+b_u+b_i$ is the baseline estimation of the rating given the user and the item. It does not include anything about the general preferences of the user or the features of the item. Simply, it is the most naive estimate given the user and the item.

Before explaining the rest of the formula, we must introduce two new concepts: implicit and explicit feedback. Explicit feedback is when users state how much they like/prefer an item explicitly. There are no reasons to further analyze them as they are strong statements in themselves. However, they can also give implicit feedback. For example, a user might decide to watch a movie, but might not rate it or a user can watch only a part of the movie. This kind of information is generally not completely descriptive in itself and have to be interpreted. For instance, number of minutes watched does not mean the same thing for two different length videos and this must be taken into account to get a better understanding of preferences of a user. Currently, implicit feedback is used even more than explicit feedback as generally people tend to produce more implicit feedback than the explicit ones.

Normally, it seems like we only have explicit feedback given our data set. However, it can also be argued that we still have some kind of implicit feedback: the fact that the users have rated these movies show that they watched them. Even this is useful for the recommendation system. For example, if the user has not watched 'Star Wars', even if the rest of the data shows the user likes Sci-fi movies, it might be better to recommend a different movie than 'Empire Strikes Back' as the user implicitly showed less interest for the 'Star Wars Saga'. Thus, in our case, the implicit feedback is '1' if the user has rated the movie and it is '0' if he or she has not. The set of items which the user has given explicit feedback about is stated as $R(u)$ while the set of items which the user has given implicit feedback about is stated as $N(u)$. Since we get our implicit feedbacks from the explicit feedbacks, these two sets are same in our implementation.

We may now return to the formula. The value after the baseline is the part where the effect of latent factors comes into play. It can be restated as:
$$q_i^T\left(p_u + \abs{N(u)}^{-1/2}\sum\nolimits_{j\in N(u)}y_j\right)$$
Here, $q_i$ and $p_u$ are two size $K$ vectors which are the representations of the user $u$ and the item $i$ in the factor space. These are acquired using the explicit feedback matrix, and so $p_u$ only shows the explicit preferences of the user. On the other hand, $y_j$ are again size $K$ vectors which can be thought as representations of the implicit feedback about the item $j$ in the factor space. If we follow the example again, not having watched 'Star Wars' might mean that the user might not be as into Sci-fi as the explicit data states. Then, the feature that represents this preference should be lower than someone who has watched all of this series (if this feature exists). All effects from the implicit feedback are added together to see the total effect of implicit feedback on $p_u$. So, $p_u + \abs{N(u)}^{-1/2}\sum_{j\in N(u)}y_j$, can be thought as a preference vector for the user combining both implicit and explicit feedback. Thus, inner product is simply the expected rating given features of the item and preferences of the user. The last two values are the part where neighbor approach has an effect. One difference from the first part is that the filtering is item based. So, we find the neighbors of items rather than users. It can be restated again as:
$$\abs{R^k(i,u)}^{-1/2}\sum_{t\in R^k(i;u)}(r_{ut}-b_{ut})w_{it}+\abs{N^k(i,u)}^{-1/2}\sum_{t\in N^k(i;u)}c_{it}$$
As in the first part, $k$ represents the number of neighbors of the item $i$. $R^k(i,u)$ is the set of items which are neighbors of $i$ that $u$ has given explicit feedback about. $N^k(i,u)$ is its implicit counterpart and for us those two sets are same. $\sum_{t\in R^k(i;u)}(r_{ut}-b_{ut})w_{it}$ is the weighted sum of the deviation of the ratings of item $i$'s neighbors from the baseline estimation. The weights $w_{it}$ represent the similarity between the item $i$ and $t$. If $i$ is similar to $t$ and the user has rated $t$ different than expected, then we expect the user to rate $i$ similarly different. This is the effect of explicit feedback. The second term, $\sum_{t\in N^k(i;u)}c_{it}$, is the sum of the effects of implicit feedback given about the neighbors of $i$. Thus again, the effects of implicit and explicit feedback about the neighbors are combined to have a total effect. 

In the end, this estimation method combines the best of two worlds. The latent factors part captures the general effect of the set: what the user generally prefers and what the item can be classed as. Only including this part however misses the specific effects. On the other hand, the neighborhood approach looks at the specific items more than the general effects, which might give good estimates for very similar movies, but lose some general information about the set since it only uses some of the information about the user. Combining these results into a method that includes both general and local effects give the most accurate algorithm to estimate the ratings.

\subsubsection{Implementation}
In the estimation formula at \eqref{1}, we only have the values $\mu$, $r_{ut}$ and $b_{ut}$ values deterministically acquired from the dataset. The rest of the parameters ($b_u, b_i, q_i, p_u, y_j, w_{it}, c_{it}$) and the k-neighborhood sets have to be found using the training data. So, we start the algorithm by estimating the k-neighborhood sets as they will be used for estimating the parameters as well. These are found in a similar way as the first method; by using a similarity metric (pearson correlation coefficient in this case). However, this time we also regularize these values to avoid overfitting. This is done as:
$$s_{ij} = \frac{n_{ij}}{n_{ij}+\lambda_1}\rho_{ij}$$
where $\rho_{ij}$ is the pearson coefficient and $n_{ij}$ is the number of users that have rated both $i$ and $j$. $\lambda_1$ is a regularizing parameter that has to be optimized for the best result. These $s$ values are used to find the neighborhoods and also used for initializing the $w_{ij}$ and $c_{ij}$ values. Increasing $k$ (the number of neighbors) generally increases the accuracy of the estimation, but it also increases the running time. After finding the $k$-neighborhood sets, we also find the initial values for the remaining parameters. The baseline values are initialized simply as the average rating of the user $u$ minus the total average for $b_u$ and the average rating for the item $i$ minus the total average for $b_i$. Values $p_u$ and $q_i$ are initially found by using SVD to make a rank-$K$ approximation, similar to the second method. $K$ value here is also a parameter to be optimized to get better results. Values $y_j$ are initialized as zero since we do not have much information about them. We then continue on the actual training part of the implementation.

\begin{figure}[t]
  \centering
    \includegraphics[height=0.25\textwidth]{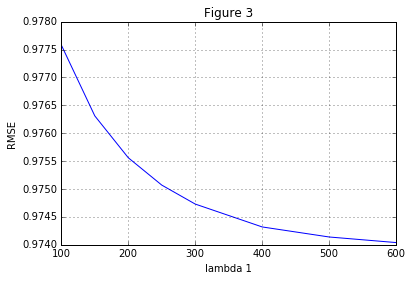}
    \caption{$RMSE$ vs $\lambda_1$ when $k=150$ and $K=10$}
\end{figure}
In the training part, we follow the reasoning in \cite{koren2008factorization} and use stochastic gradient descent to find the variables by minimizing the regularized squared error. In each iteration of the algorithm, the estimation error is calculated as $e_{ui}=r_{ui}-\hat{r}_{ui}$ for each  $(user, item, rating)$ trio and the parameters are updated using the gradient of each parameter as stated below:
\begin{align*}
    b_u \leftarrow& b_u + \gamma_1(e_{ui}-\lambda_2\cdot b_u)\\
    b_i \leftarrow& b_i + \gamma_1(e_{ui}-\lambda_2\cdot b_i)\\
    q_i \leftarrow& q_i + \gamma_2\left(e_{ui}\left(p_u+
    \abs{N(u)}^{-1/2}\sum\limits_{j\in N(u)}y_j\right)-\lambda_3\cdot q_i\right)\\
    p_u \leftarrow& p_u + \gamma_2(e_{ui}\cdot q_i - \lambda_3\cdot q_i \\
    \forall j \in &N(u):\\
    y_j \leftarrow& y_j + \gamma_2(e_{ui}\cdot\abs{N(u)}^{-1/2}\cdot q_i - \lambda_3 \cdot y_j)\\
    \forall t \in &R^k(i,u):\\
    w_{it} \leftarrow& w_{it} + \gamma_3\left(\abs{R^k(i,u)}^{-1/2}\cdot e_{ui}(r_{ut}-b_{ut})-\lambda_4\cdot w_{it}\right)\\
    \forall t \in &N^k(i,u):\\
    c_{it} \leftarrow& c_{it} + \gamma_3(\abs{R^k(i,u)}^{-1/2}\cdot e_{ui}-\lambda_4\cdot c_{it})
\end{align*}

The order of these $(user, item, rating)$ trios are arranged randomly in each iteration to prevent the deterministic order from descending into a local minimum. The algorithm is run for 6 iterations to limit the running time and the resulting parameters are used for the estimation.

In the implementation there are nine meta-parameters ($\lambda_1, \lambda_2, \lambda_3, \lambda_4, \gamma_1, \gamma_2, \gamma_3, k, K$) that has to be found using cross validation. However, since using grid search to find all these values would take too much time, we have limited the number of parameters to optimize. We initialize them with the values in \cite{koren2008factorization} and changed them one at a time to see which ones have greater effect. We found out that $\lambda_1, k$ and $K$ are the most important parameters and decided to only optimize them while leaving the rest as the values in \cite{koren2008factorization} ($\lambda_2 = 0.005, \lambda_3 = 0.015, \lambda_4 = 0.015, \gamma_1 = 0.007, \gamma_2 = 0.007, \gamma_3 = 0.001$). Also as suggested, the $\gamma$ values are reduced by a factor of 0.9.

To find the optimal values for $\lambda_1, k$ and $K$, we decide to optimize them like coordinate gradient descent: minimizing $k$ first, then using this $k$, minimizing $\lambda_1$ and at last minimizing $K$ by using the other two. The accuracy is estimated by the average $RMSE$ that is acquired by cross validation. The first result is that increasing k always increases the accuracy as stated. In the end, we chose $k = 300$ not to increase the running time further. Then, we also find out that increasing $\lambda_1$ also increases the accuracy constantly, but it converges after a while. This result can be seen in Fig. 3. Thus, we choose $\lambda_1 = 600$ as it is the largest value we have tried and it converged enough to stop trying more. Lastly, we tune $K$ to get the last of our parameters, which can be seen in Fig. 4. The last value of K is set to 10.

\begin{figure}
  \centering
    \includegraphics[height=0.25\textwidth]{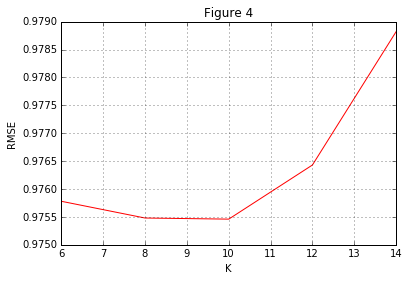}
    \caption{$RMSE$ vs $K$ when $k=150$ and $\lambda_1=200$}
\end{figure}

\section{Conclusion}

To sum up, we evaluate three different algorithms to predict the missing ratings of a given sparse user-movie rating matrix. The first two methods are simple implementations of neighborhood and latent factor approaches, and both of these methods give us decent results. From the $RMSE$ results on the test set, we see that simple matrix factorization gives better results compared to user-based collaborative filtering. Then, we present the last method, which mainly integrates those two approaches. Our final result leads to an $RMSE$ of 0.97075 which is the best $RMSE$ among the methods we try. Still, there can be further improvement if all of the meta parameters are optimized using grid search and cross validation. In the end, the integrated method is the best one according to our evaluation metric but simple matrix factorization still gives a good $RMSE$ when its simplicity is considered.

\begin{table}
\caption{$RMSE$ RESULTS OF THE METHODS ON TEST SET WITH TRAINING ALL OF THE GIVEN RATINGS}
\begin{tabular}{|*{5}{c|}}
    \hline
    Methods & User-based CF & Iterative Matrix Factorization & Integrated Model \\ \hline
    Result & \ 0.99802 & \ 0.98893 & \ 0.97075 \\
    \hline
\end{tabular}
\end{table}

\ifCLASSOPTIONcaptionsoff
  \newpage
\fi
\balance
\bibliographystyle{IEEEtran}  
\bibliography{references}

\end{document}